A New Fast Weighted All-pairs Shortest Path Search Algorithm
Based on Pruning by Shortest Path Trees

Yasuo Yamane[1] and Kenichi Kobayashi[2]
[1] Fujitsu Limited, 17-25, Shinkamata 1-chome, Ota-ku, Tokyo, 144-8588, Japan
[2] Fujitsu Laboratories Ltd., 4-1-1, Kamikodanaka, Nakahara-ku, Kawasaki, 211-8588, Japan

Abstract

Recently we submitted a paper whose title is “A New Fast Unweighted All-pairs Shortest Path Search Algorithm Based on Pruning by Shortest Path Trees” to arXiv. This is related to unweighted graphs.

This paper also presents a new fast all-pairs shortest path algorithm for weighted graph based on the same idea. In Dijkstra’s algorithm which is said to be fast in weighted graphs, the average number of accesses to adjacent vertices (expressed by α) is about equal to the average degree of the graph. On the other hand, our algorithm utilizes the shortest path trees of adjacent vertices of each source vertex in the same manner as the algorithm for unweighted graphs, and reduce α drastically in comparison with Dijkstra’s algorithm. Roughly speaking α is reduced to the value close to 1, because the average degree of a tree is about 2, and one is used to come in and the other is used to go out, although that does not hold true when the depth of the short path trees is small.

In case of weighted graphs, a problem which does not occur in unweighted graphs occurs. It is waiting for the generation of the shortest path tree of an adjacent vertex. Therefore, it is possible that a deadlock occurs. We prove our algorithm is deadlock-free.

We compared our algorithm with Dijkstra’s and Peng’s algorithms. On Dijkstra’s ours outperforms it on speed and α except that Dijkstra’s slightly outperforms ours or they are almost the same on CPU time in sparse scale-free graphs. The result on Peng’s is as follows: In speed and α, ours outperforms Peng’s in hypercube-shaped and dense scale-free graphs, but conversely Peng’s outperforms ours in sparse scale-free graphs.

## 1. Introduction

Recently we submitted a paper [Yamane19], which proposes a fast unweighted all-pairs shortest path search algorithm called PST. The feature is pruning by shortest path trees. We call it PSTu (Pruning by Shortest path Tree for unweighted graphs) in this paper.

In this paper, we present a similar algorithm for weighted graphs, which we call PSTw below to distinguish from PSTu.

All-pairs shortest path algorithms have many applications in general graphs, for example, railroad networks, transportation networks, Web, and SNS (Social Networking Service).

Dijkstra's algorithm is well-known as a fast algorithm to compute the shortest paths for weighed graphs. Dijkstra's algorithm, accesses all the adjacent vertices at each vertex, so the average number of access times to adjacent vertices at each vertex, which is expressed by $\alpha$ below, is about equal to the average degree of the graph. On the other hand, like PSTu, PSTw utilizes the shortest path trees of adjacent vertices of each source (starting) vertex when traversing from the source vertex beyond the adjacent vertices, and reduce $\alpha$ drastically in comparison with Dijkstra's algorithm. Roughly speaking, $\alpha$ of PSTw is reduced to the value close to 1, because the average degree of a tree is about 2, and one is used to come in and the other is used to go out, although that does not hold true when the depth of the short path trees is small as mentioned later.

In case of weighted graphs, a problem which does not occur in unweighted graphs occurs. It is waiting for the generation of the shortest path tree of an adjacent vertex. Therefore, it is possible that a deadlock occurs. We prove our algorithm is deadlock-free.

We compared PSTw with two fast algorithms. One is one of the representative algorithms [Dijkstra] where Dijkstra's algorithm is applied for computing the all-pairs shortest paths. We call it simply Dijkstra below. The other is Peng's algorithm [Peng12] and we call it Peng simply below. Peng is said to be one of the sate-of-the-art algorithms in [Kim18], and $\alpha$ of it is amazingly small in some case in our experiments.

PSTw outperforms Dijkstra on speed and $\alpha$ except that Dijkstra's slightly outperforms PSTw or they are almost the same on CPU time in sparse scale-free graphs. The result of comparison with Peng is as follows: On speed and $\alpha$, PSTw outperforms Peng in hypercube-shaped and dense scale-free graphs although Peng outperforms ours in sparse scale-free graphs.

## 2. Relative Works

In this section we explain the conventional algorithms for weighted all-pairs shortest path search.

1) Floyd-Warshall algorithm ([Floyd62], [Warshall62])

It is one of the most famous algorithms for all-pairs shortest algorithms. This is an algorithm for weighted graphs, but It can be applied to unweighted graphs letting all weights be one. The (worst) time complexity is $O(n^3)$ where $n$ is the number of vertices, but the implementation is very simple requiring only some lines. Let $n$ and $m$ mean the numbers of vertices and edges of a graph respectively, and time complexity mean worst one below.

2) Dijkstra's algorithm ([Dijkstra])

It is well-known as a fast algorithm for computing the shortest paths from a source vertex to the all vertices in weighted graphs, and the time complexity is $O(n\ log\ n + m)$. It can be applied for computing the all-pairs shortest paths by letting each vertex as a source one, and the time complexity is $O\big(n(n\ log\ n + m)\big)$. To distinguish them, we call the former SS-Dijkstra (Single-Source Dijkstra), and the latter AP-Dijkstra (All-Pairs Dijkstra) or simply Dijkstra.

3) Peng’s algorithm ([Peng12])

This is a variant of AP-Dijkstra, and it computes the shortest paths from each vertex to all the vertices serially. The feature of it is to utilize the length of the already computed shortest paths to reduce $\alpha$.

4) breadth-first search ([BFS])

This algorithm is used for unweighted graphs, but explained because it is used below. Breadth-first search is originally an algorithm to traverse all the vertices in breadth-first manner, and it is applied for various purposes. The time complexity is $O(n + m)$. It can be applied for computing the all-pairs shortest paths by letting each vertex as a source one, and the time complexity is $O\left(n(n + m)\right)$. To distinguish from an original breadth-first search, we call this algorithm AP-BFS (All-Pairs shortest path Breadth-First Search) or simply BFS.

## 3. PSTw Algorithm

At first, we explain PSTu simply. PSTu is a variant of AP-BFS. They are largely different in the following two points:

1) Pruning by shortest path trees

   AP-BFS accesses the all adjacent vertices at each vertex. On the other hand, PSTu reduces $\alpha$ to the value to close to 1 drastically roughly speaking, although that does not hold true when the depth of the short path trees is small.

2) Synchronous generation of shortest path trees

   AP-BFS applies breadth-first search to each vertex and create shortest path trees serially. On the other hand, PSTu create shortest path trees synchronously at each vertex like creating concentric circles in increasing order.

PSTw is likewise a variant of AP-Dijkstra. The relationship between PSTw and AP-Dijkstra is quite the same as that between PSTu and AP-BFS. In AP-Dijkstra, SS-Dijkstra is serially applied at each vertex like AP-BFS. And PSTw is largely different from AP-Dijkstra in the two points mentioned above.

Although PSTw and PSTu are similar, there are some differences such as waiting, which might cause a deadlock problem.

In 3.1, the differences from PSTu are mainly mentioned. In 3.2, the details of PSTw are explained. And in 3.3, it is proved that PSTw is deadlock-free.

### 3.1 The differences from PSTu

First of all, PSTu is explained, on which PSTw is based. Generally, in all-pairs shortest path algorithms, a shortest path tree is generated at each vertex. It is well-known that the shortest paths from each vertex $v$ to all the vertices can be compactly expressed by a tree whose root is $v$. This tree is called a "shortest path tree" and represented by $T(v)$ below. Fig.3.1(b) shows the shortest path tree for a weighted graph shown in (a). PSTu utilizes shortest path trees being generated for pruning.

When vertices $v$ and $w$ is adjacent, $T(v)$ and $T(w)$ is very similar. In addition, as mentioned in [Yamane19], $T(w)$ has only to be traversed when traversing through $w$ from $v$. In other words, it is unnecessary to traverse the edges which are not contained in $T(w)$. Based on this idea, PSTu is a variant of AP-BFS to which the following two modifications are added.

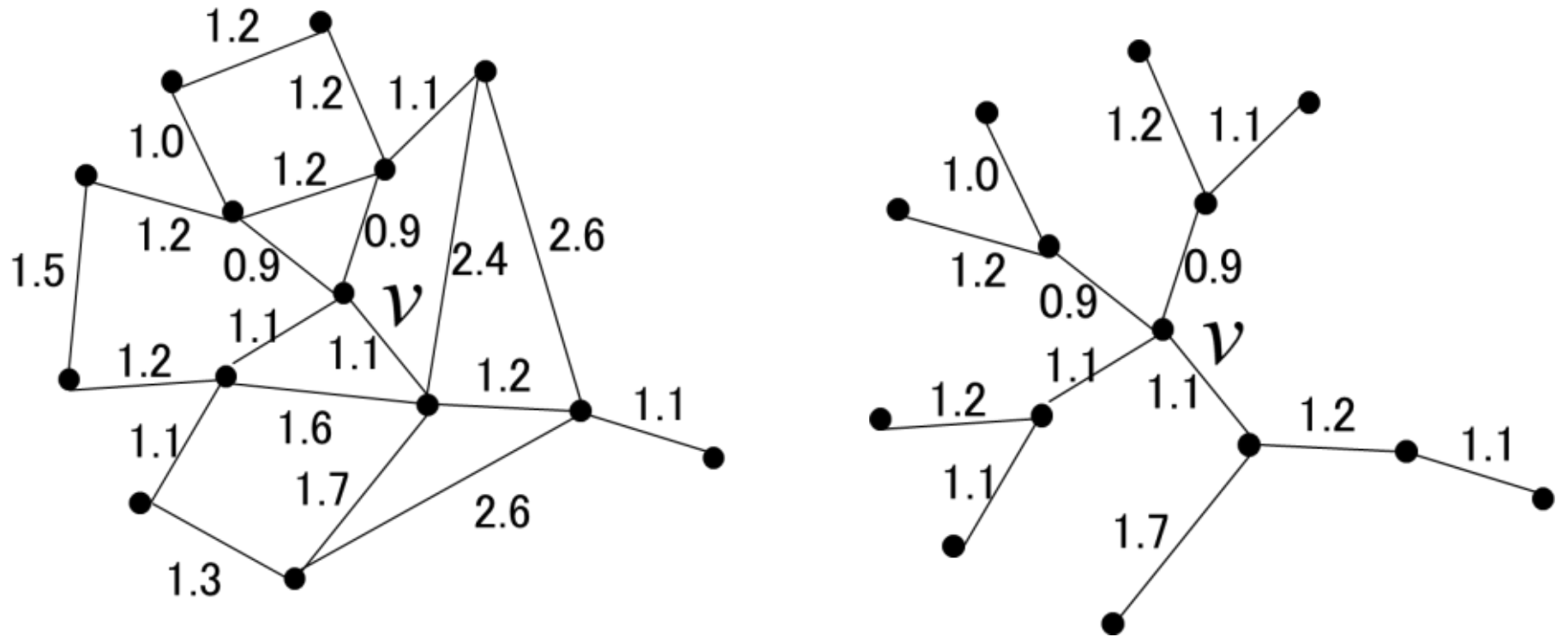

(a) a weighted graph (b) a shortest path tree

Fig.3.1 A weighted graph and a shortest path tree

1) Pruning by shortest path trees of adjacent vertices

Let us consider two adjacent vertices $v$ and $w$. When searching through $w$ from a source vertex $v$, only $T(w)$ , the shortest paths to all other vertices from $w$, is traversed. Roughly speaking this modification reduces $\alpha$ drastically in comparison with Dijkstra's algorithm to the value close to one because $\alpha$ of a tree is generally about two and one of them is used to go in although this does not hold true when the depth of shortest path trees is small. Here, it seems worth to note the accesses to adjacent vertices from a source vertex and those from adjacent vertices to their adjacent vertices cannot be reduced to understand that it does not hold true.

2) Synchronous generation of shortest paths trees

In AP-BFS, breadth-first search is done at each vertex serially. This minimize the usage of memory. However, in PSTu it is necessary to create shortest path trees synchronously to realize 1). Concretely speaking, at each vertex $v$, the parts of $T(v)$ is synchronously generated, and they are like generating concentric circles in increasing order of radius. First the part to the vertices adjacent to $v$, that is, the vertices on the circle centering around $v$ whose radius is 1 is generated synchronously with all the other vertices. Secondly the part to the vertices on the circle whose radius is 2 is generated synchronously as if the computation is done in parallel, and so on.

The above is a simple explanation of PSTu. PSTw is a variant of PSTu where the idea mentioned above is applied to weighted graphs, and has the same features mentioned in 1) and 2) above. However, the method to realize 2) in PSTw is different, which we explain below.

It is the same as PSTu to traverse $T(w)$ when searching through $w$ adjacent to a source vertex $v$, and to create $T(v)$ synchronizing with $T(w)$ and other vertices because it is impossible to traverse $T(w)$ if the necessary parts of it is not created. The largely different thing is as follows: In case of unweighted graphs, it is possible to synchronize like concentric circles. This ensures that the necessary parts of $T(w)$ is created when searching through $w$ from $v$. However, in case of weighted graphs, it cannot be ensured because the weights of the edges are various.

And so letting the sequence of all vertices be $v_1, v_2, \cdots, v_n$, PSTw synchronizes by searching little by little from each source vertex $v_i$ in this order with $v_1$ after $v_n$. However, it is not possible to synchronize perfectly like PSTu, and there are such cases as when searching $T(w)$ through $w$ adjacent to $v_i$ and trying to search beyond some vertex $x$ of $T(w)$, the children of $x$ are not created yet. Then PSTw suspends searching from $v_i$, and continues searching from the next vertex of $v_i$, and waits for the children of $x$ to be created.

3.2 The details of PSTw

PSTw is a variant of AP-BFS to which the following two modifications are added to realize 1) and 2) mentioned above. We implemented for undirected graphs but it is easy to modify for directed graphs. We add notice when necessary.

1) Modification for pruning by shortest path trees

This is modification of the data structure representing shortest path trees. In AP-Dijkstra, $T(v)$ of each source vertex $v$ can be represented by a vector of size n. That is, letting all the vertices be $v_1, v_2, \cdots, v_n$ and the vector be $[s_i]$, it is sufficient to let $s_i$ be $j$ of the parent $v_j$ of $v_i$ on $T(v)$. All the shortest path trees can be represented by an $n$ by $n$ matrix. However, in PSTw, it is necessary to memorize children in shortest path trees to utilize pruning by shortest path trees. So, for each vertex, a shortest path tree retaining the relationship between the parents and the children is generated.

2) Modification for synchronous generation of shortest paths trees

Priority queues are used in the same manner as AP-Dijkstra. And when searching from a source vertex $v_i$, the first vertex $x$ is dequeued from a priority queue $q$ for $v_i$, and the children of $x$ on $T(w)$ are tried to traverse. However in case that they are not created, searching from $v_i$ is suspended, and $x$ is enqueued to $q$ again, and searching from the next vertex of $v_i$ is continued.

The following two classes are shown as the data structures for PSTw based on the modifications. One class is $Vertex$ which represents a vertex $v$ on a graph, and the other is $T_Vertex$ which represents a vertex on the shortest path tree $T(v)$. To distinguish vertices of two classes, we call the vertex of $Vertex$ a vertex simply, and that of $T_Vertex$ a "t-vertex".

A t-vertex is different from a vertex in such a point as it has a parent and children. Letting a vertex be $x$, we express the t-vertex corresponding to $x$ by $x'$ below. We say "$x'$ corresponds to $x$" when the identifies of them are identical.

1) $Vertex$

An instance of this class consists of five properties $id$, $adj_vertices$, $root$, $c$ $que$, and $t_v_hash$. Each vertex $v_i$ $(1 \le i \le n)$ has an id whose value is $i$, and property $id$ represents $i$. Let this instance represent a vertex $v$. Property $adj_vertices$ represents the set of a pair $(w,\ e)$ where $w$ is adjacent to $v$ and $e$ is the length of edge from $v$ to $w$. The class of $w$ is $Vertex$. Property $root$ represents a t-vertex $v'$ in $T(v)$, that is, the root of $T(v)$. Property $que$ represents a priority queue prepared for each vertex explained above. Although we do not explain the class of a priority queue in detail, but each queue $q$ has the following four methods:

$q.enque(x, d)$ enqueue a vertex $x$ whose distance from $v$ is $d$

$q.deque()$ return a pair of $x$ and $d$ at the top of $q$, and $None$ when $q$ is empty

$que.update(x', d, d_{new})$ update the $d$ of already enqueued pair $(x', d)$ to $d_{new}$

$que.len()$ return the number of enqueued elements.

We implemented the priority queue using python's list and by inserting an enqueued element into the appropriate position although heap is said to be the fastest. Peng and Dijkstra were implemented likewise.

The following summarizes what is mentioned above.

| | |
|---|---|
| $id$ | the id of $v$ from 1 to $n$ |
| $adj_veritces$ | the set of a pair of $(w,\ e)$ where $w$ is adjacent to $v$ and $e$ is the length of the edge from $v$ to $w$ |
| $root$ | the root of $T(v)$ whose class is $T_Vertex$ |
| $que$ | a priority queue for $v$ |
| $t_v_hash$ | a hash table whose key is id and value is t-vertex |

2) $T_Vertex$

An instance of this class consists of six properties $vertex$, $cor$, $parent$,

$children$, $is_determined$ and $parent_edge_len$. $T$ of $T_Vertex$ means (shortest path) tree. Let this instance represent a t-vertex $x'$ on $T(v)$, and $x'$ be reached through a t-vertex $w'$ where $w$ and a source vertex $v$ are adjacent. Then the property $vertex$ represents $x$. The property $cor$ represents the t-vertex corresponding to $x'$ on $T(w)$. This t-vertex is represented by $x''$ below. That is, the value of $cor$ is $x''$. The property $parent$ represents the parent of $x'$ in $T(v)$. The property children represents the set of the children of $x'$ in $T(v)$. The property $is_determined$ shows whether the shortest path from $v$ to $x$ is determined. If it is determined, the value of the property is true. When the value of the property $is_determined$ of $x''$ of $T(w)$ is true, that means the children of $x''$ is also created so it is unnecessary to wait for traversing them. The property $parent_edge_len$ represents the edge length from $x'$ to its parent on $T(v)$.

The values of properties of $x'$ and these relationships between the data structures explained above are summarized in Fig.3.2.

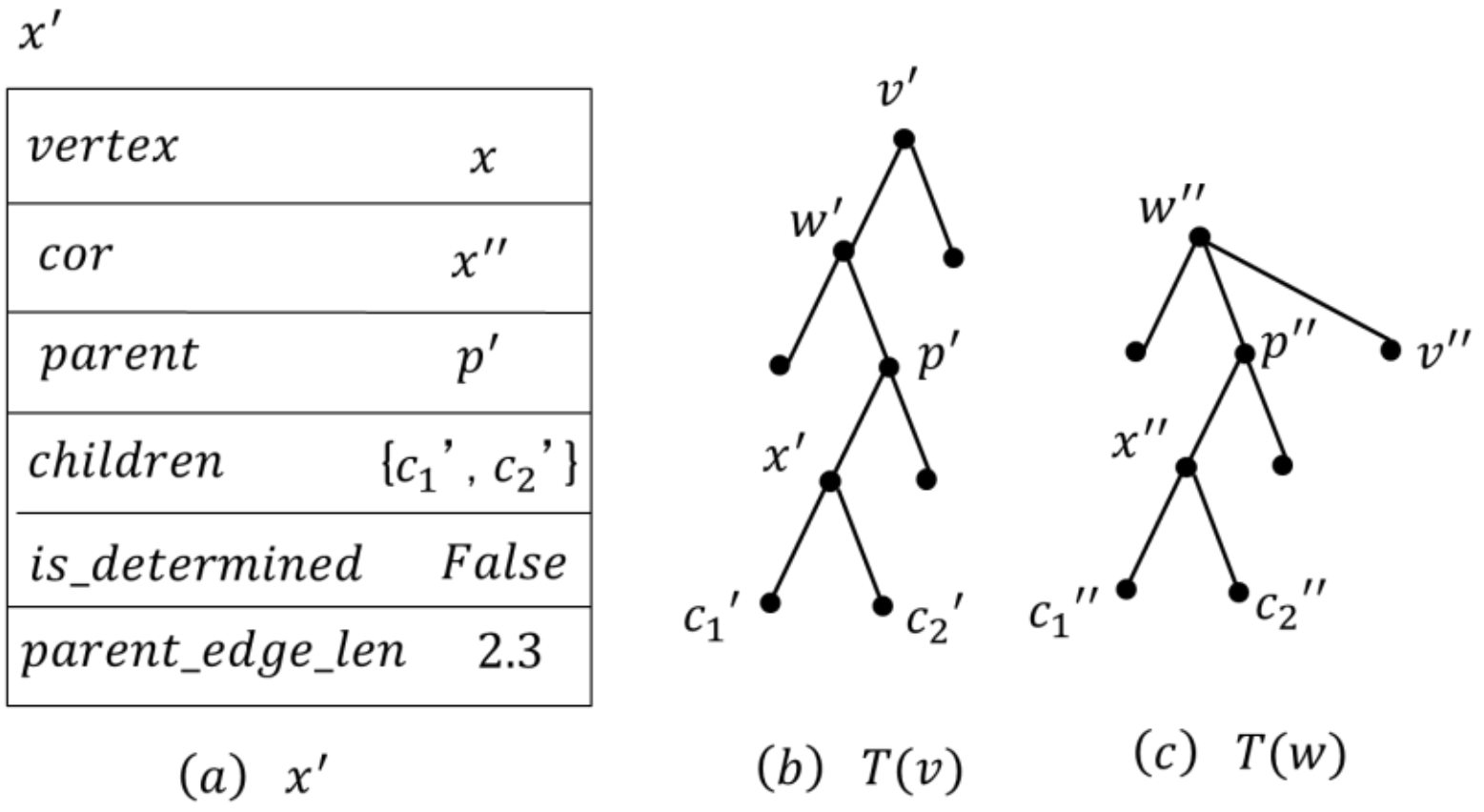

Fig3.2 the data structure of a t-vertex and the relationship between t-vertices and shortest path trees

The python-like pseudo code of the algorithm for creating a graph $G = (V, E)$ where $|V| = n$ and initialization is shown, where the definition of classes $Vertex$ and $T_Vertex()$ are also shown.

In python's class definition, the name of a method creating a new instance is $__init__$, and the first parameter of it is $self$ which represents a new instance. However, we use an understandable parameter name instead of $self$, for example $v$.

We implemented the creation of a graph and initialization together in our experiments, but they should be separated to be exact because it relates to the measurements as mentioned later. Here, initialization means preparation for PSTw algorithm; for example, initialization of a queue at each vertex.

---

Algorithm for creation of a graph and initialization

input: $n$

output: $G = (V, E)$

---

class $Vertex$():

  $_init_(v, id)$:   # returns a new instance $v$

    $v.id = id$

    $v.adj_vertices = \{\ \}$

    $v.root = None$

    set an empty priority queue to $v.que$

    set an empty hash table to $v.t_v_hash$

set an empty set to $V$

$for\ id \in \{1, 2, \cdots, n\}$:

  $v = Vertex(id)$   # create a new instance of $Vertex$

  add $v$ to $V$

$for\ v \in V$:

  $for\ w \in a\ set\ of\ v's\ adjacent\ vertices$:

    $edge_length$ = the length of edge from $v$ to $w$

    add $(w, edge_length)$ to $v.adj_vertices.$

---

. Secondly the main routine of PSTw algorithm, and the definition of function $extend(\ v, D, S)$, which extends $T(v)$, are shown. The function returns $True$ if the function need to be called again on $v$, and otherwise $False$. The definition of classes $T_Vertex()$ is also shown. $D$ is an $n$ by $n$ matrix to represent distances between vertices. It is called a "distance matrix". Its element $D[i, j]$ means the distance from $v_i$ to $v_j$. All the elements of $D$ are initialized to 0.0. $S$ is an $n$ by $n$ matrix to represent all the generated shortest path trees. It is called a "shortest path tree matrix". Its element $S[i, j]$ represents the parent of $v_i$ on the shortest path from $v_i$ to a source vertex $v_j$. The diagonal elements of $S$ are initialized to $NO_PARENET$ $(= -1)$, and all the other elements to $NOT_SEARCHED$ $(= 0)$. $NO_PARENET$ means that there is no parent, that is, $v_i$ is a source vertex. NOT_SEARCHED means that $v_i$ has not been searched yet.

It is worth noting that although at first $V$ represents the set of all the vertices, $V$ virtually represents the set of vertices whose all the shortest paths form source vertices have not been computed yet.

---

PSTw algorithm

  input: $G = (V, E)$

  output: $D$ and $S$

---

```
class T_Vertex():
    _init_(v', v, v'', p', e):      # returns a new instance v'
        v'.vertex = v
        v'.cor = v''
        v'.parent = p'
        if p' is not None:
            add v' to p'.childeren
            v'.parent = p'
        else:
            v'.parent = None
        set an empty set to v'.children
        v'.is_determined = False
        v'.parent_edge_len = e

set an n by n initialized matrix to D
set an n by n initialized matrix to S
while 0 < |V|:
    set an empty set to V_new
    for v ∈ V:
        if extend(v, D, S)
            add v to V_new
    V = V_new

def extend(v, D, S):
    r = True
    if len(v.t_v_hash) == 0:
        v.root = v' = T_Vertex(v, None, None, None)      # create a new instance of T_Vertex
        v'.is_determined = True
        v.t_v_hash[v.id] = v'
    elif len(v.t_v_hash) == 1:
        for (w, e) ∈ v.adj_vertices:
            D[w.id, v.id] = edge_len
            S[w.id, v.id] = v.id
            w' = T_Vertex(w, w.root, v.root, e)          # create a new instance of T_Vertex
            v.que.enque(w', e)
            v.t_v_hash[w.id] = w'
```

```
else:
    (w', e) = v.que.deque()
    if not w''.is_determined:
        v.que.endqu(w', edge_len)      # waiting                ---- (*)
    else:
        w'.is_determined = True
        for x'' in w''.children:
            x = x'.vertex
            if x.id = v.id: continue
            w = w'.vertex
            e = x''.parent_edge_len
            d = D[w.id, v.id ] + e
            x = x'.vertex
            if S[x.id, v.id] == NOT_SEARCHED:
                x' = T_Vertex(x, x'', w', e)          # create a new instance of T_Vertex
                v.que.enque(x', d)
                v.t_v_hash[x.id ] = x'
                D[x.id, v.id] = d
                S[x.id, v.id] = w.id
            else:
                x' = v.t_v_hash[x.id ]
                if d < D[x.id, v.id]:
                    v.que.update(D[x.id , v.id], x', d)
                    x'.cor = x''
                    delete x' from x'.parent.children
                    x'.parent = w'
                    add x' to w'.children
                    x'.parent_edge_len = e
                    D[x.id, v.id] = d
                    S[x.id, v.id] = w.id
    if v.que.len( ) == 0: r = False
return r
```

In PSTu, for each vertex $v$, we used a counter for the number of vertices whose shortest paths are obtained, and aimed at stopping the processing on $v$ as soon as it becomes $n$ before a queue become empty. We also tried it in Peng but failed, so we removed it from PSTw and Dijkstra to make comparison fair. Consequently, it seems counters are hardly effective and slightly effective when $n$ is small.

### 3.3 Proof of PSTw being deadlock-free

Enqueueing the vertex again at (*) in the definition of function $extend()$ which is dequeued two lines above virtually makes waiting. So it is possible that a deadlock occurs in PSTw. The following theorem ensures that PSTw is deadlock-free. We assume that the weights of all the edges are positive.

Theorem

If the weights of all the edges are positive, PSTw is deadlock-free.

Proof:

We prove using proof by contradiction.

Let us assume that a deadlock occurs in $k$ vertices $v_1, v_2, \cdots, v_k$. Let $v_{k+j}=v_j (1 \leq j \leq 2)$ for convenience. Letting $1 \leq i \leq k$, let us assume $v_i$ is waiting for $v_{i+1}$. Then $v_i$ should be adjacent to $v_{i+1}$, and $v_i$ should have tried to search beyond some vertex $x_{i+1}$ of $T(v_{i+1})$ but the children of $x_{i+1}$ should have not been created yet, so $v_i$ should be waiting. Let the distance between $v_i$ and $v_{i+1}$, that is, the weight of the edge between $v_i$ and $v_{i+1}$ be $e_i$, and let the distance between $v_{i+1}$ and $x_{i+1}$ be $d_{i+1}$. Likewise, $v_{i+1}$ is waiting for $v_{i+2}$ at some vertex $x_{i+2}$ of $T(v_{i+2})$. Fig.3.3 shows the relationship among the vertices, and the distances. Here, letting $1 \leq j \leq k$, the following inequality holds true.

$$e_{i+1} + d_{i+2} \leq d_{i+1}$$

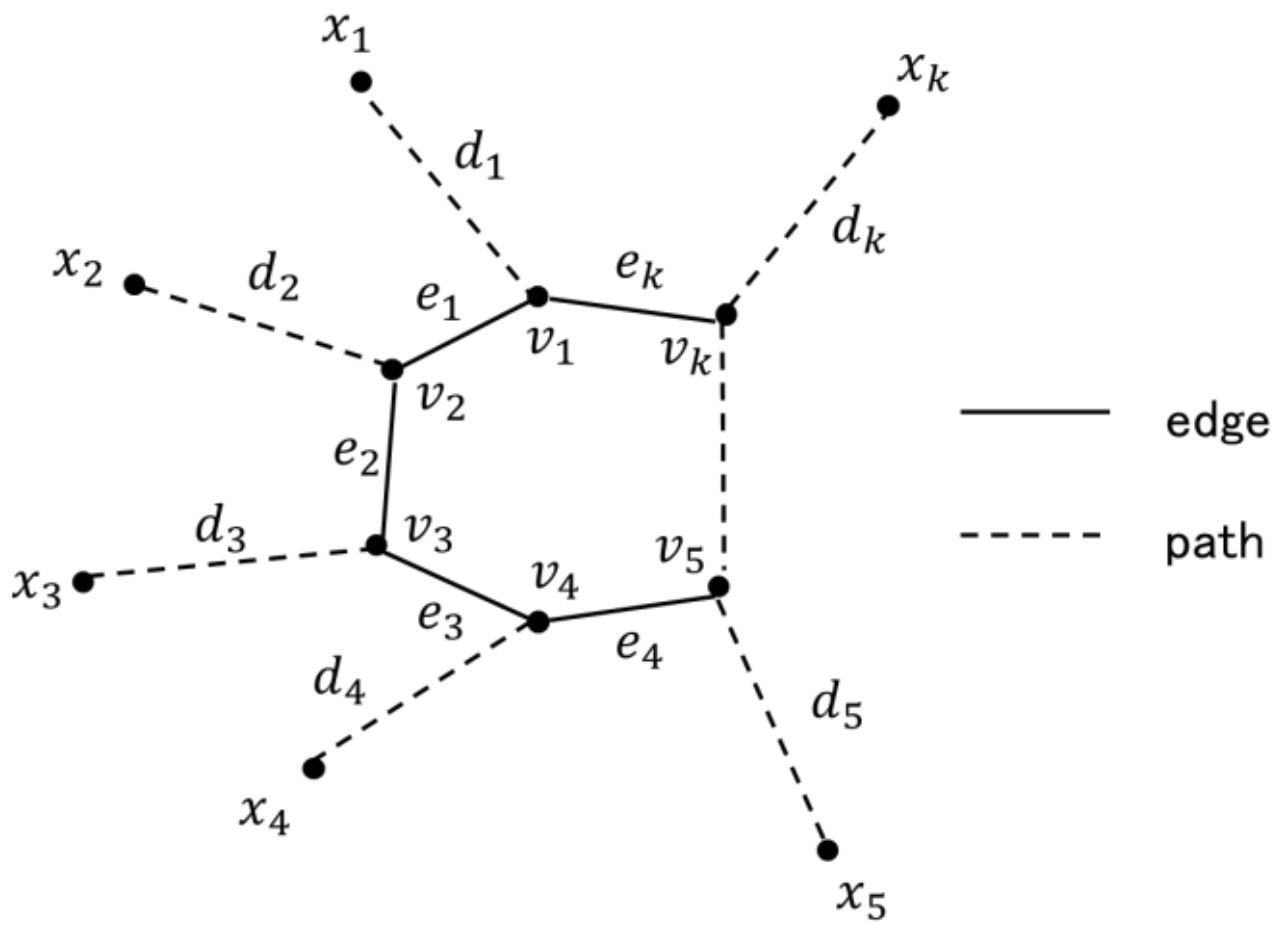

Fig.3.3 The relationship between the vertices and distances

The reason is as follows: The left-hand side of the inequality is the distance from $v_{i+1}$ to $x_{i+2}$ through $v_{i+2}$ and the right-hand side is the distance from $v_{i+1}$ to $x_{i+1}$. In searching from a source vertex $v_{i+1}$, $x_{i+2}$ should have been enqueued before $x_{i+1}$, because if not so, $v_{i+1}$ should be waiting at $x_{i+1}$. This means the distance from $v_{i+1}$ to $x_{i+2}$ should be less than or equal to the distance from $v_{i+1}$ to $x_{i+1}$, so the inequality holds true.

Therefore, the following expression can be got by summing up the both sides from $i = 1$ to $k$.

$$\sum_{i=1}^{k}(e_{i+1} + d_{i+2}) \le \sum_{i=1}^{k} d_{i+1}$$

So the following inequality holds true.

$$\sum_{i=1}^{k} e_i + \sum_{i=1}^{k} d_i \le \sum_{i=1}^{k} d_i$$

This means $\sum_{i=1}^{k} e_i \le 0$, and it contradicts the premise that the all weights are positive. Q.E.D.

4. Evaluation

We compared PSTw with AP-Dijkstra, that is, Dijkstra, and Peng in CPU time and α. We excluded Floyd-Warshall algorithm because its time complexity is $O(n^3)$ and it is worse than the time complexity of Dijkstra $O(n(n \log n + m))$. We measured letting $n = 2^{2i}(i = 3,4,5,6)$, that is, $n = 64, 256, 1024, 4096$.

On CPU time, as mentioned above, creating a graph also includes initialization. Therefore, CPU time of PSTw does not include the initialization time. We think initialization does not affect the total CPU time so much, but it should be included in PSTw to be exact. Peng and Dijkstra are also implemented in the same manner, so their CPU time also do not include initialization time.

We used the following two kinds of graphs for comparison, which are hypercube-shaped and scale-free graphs. We selected scale-free graphs because they are said to be ubiquitous in the real world. The relationship between the degree of each vertex $d$ and the frequency of the vertices whose degree is equal to $d$ obey to a power distribution. Consequently, $d$ takes various values. On the other hand, hypercube-shaped graphs have a feature quite opposite to scale-free graphs, that is, $d$ takes only one value. We selected hypercube-shaped graphs because we wanted to examine the three algorithms from two quite different viewpoints.

1) hypercube-shaped graph

It is worth to note that the degree of each vertex of this graph is $\log_2 n$, so the values of $\alpha$ are about 6, 8, 10, and 12 for $n = 64, 256, 1024$, and $4096$.

2) Scale-free graph

The graph is created as follows: When creating a graph of size $n$, first a complete graph $G = (V, E)$ of size $n'(< n)$ is created. Secondly the remaining $n - n'$ vertices are added one by one as follows: Let one of them be $v$. Let $n'$ vertices chosen randomly from $V$ be $v_1, v_2, \cdots, v_{n'}$. Then Let $V = V \cup \{v\}$ and $E = E \cup \{(v, v_1), (v, v_2), \cdots, (v, v_{n'})\}$. The probability of choosing $v_i$ $(i = 1, 2, \cdots, n')$ is let to be proportional to the degree of $v_i$. We compared in a sparse case where $n' = 2$ and a dense case where $n' = \sqrt{n}$.

For measurement environment, we used FUJITSU Workstation CELSIUS M740 with Intel Xeon E5-1603 v4 (2.80GHz) and 32GB main memory, programing language Python, and OS Linux.

## 4.1 Comparison in hypercube-shaped graphs

Fig. 4.1 and Table 4.1 show the comparison of CPU time of the two algorithms in hypercube-shaped graphs. Fig. 4.1 is shown in a double-logarithmic graph. CPU time graphs are shown in the same manner below. Table 4.1 shows the actual values in detail. Tables show the actual values likewise below. The column whose name is /PSTw shows the rate of the CPU time (called CPU time rate) of the algorithms other than PSTw against that of PSTw, that is, how many times PSTw is faster than the others. When it is smaller than 1.0, that means that PSTw is outperformed. When $n$ = 4096, PSTw is 2.5 times faster than Peng, and 2.0 times faster than Dijkstra. The CPU rates of Peng and Dijkstra tend to increase as $n$ increases.

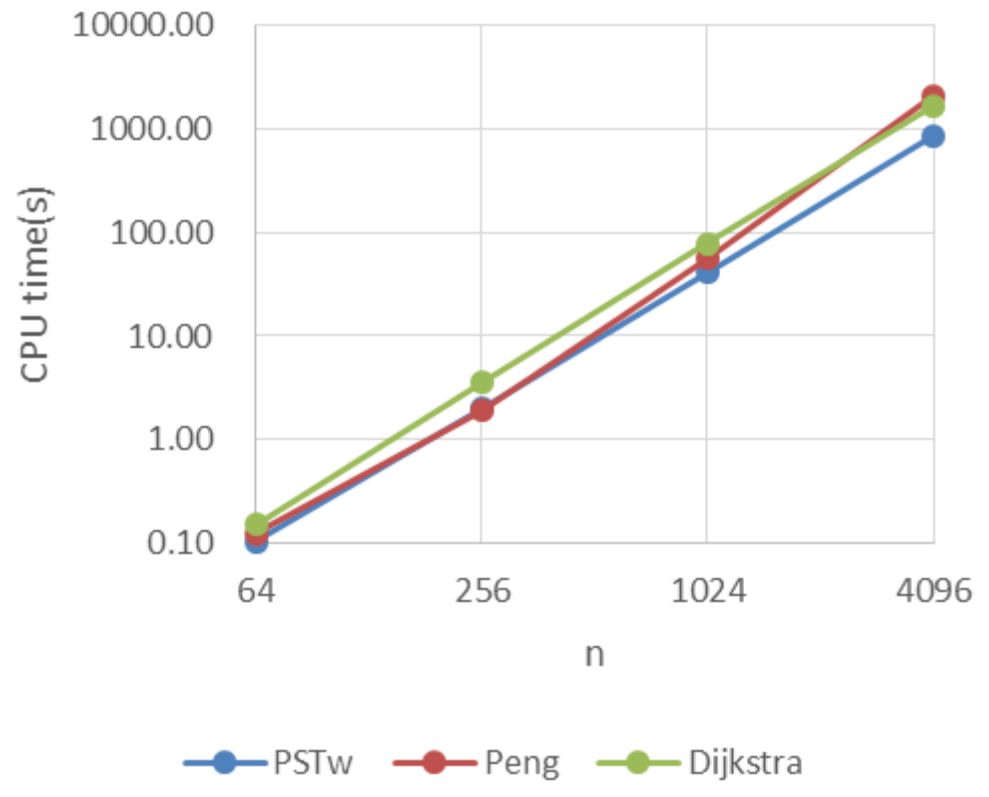

Fig. 4.1 Comparison in CPU time

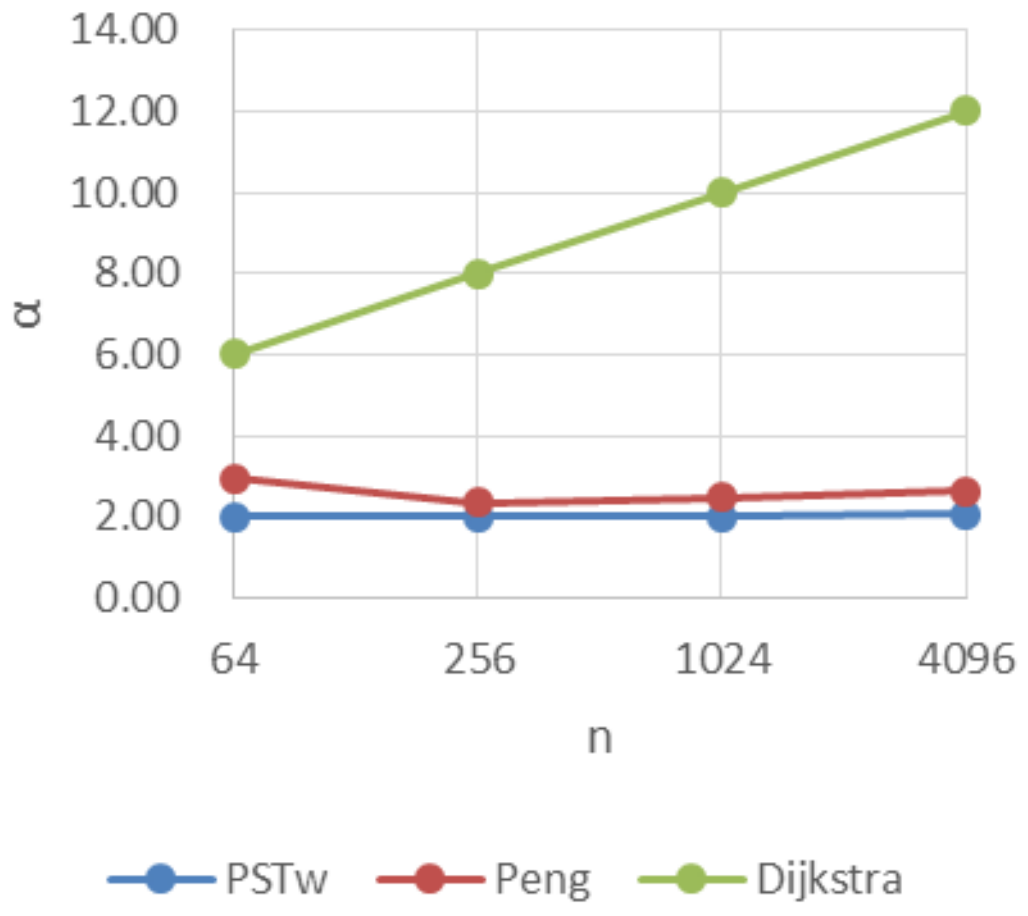

Fig. 4.2 Comparison in α

Table 4.1 Comparison in CPU time

| n | PSTw | Peng | | Dijkstra | |
|---|---|---|---|---|---|
| | time (s) | time (s) | /.PSTw | time (s) | /PSTw |
| 64 | 0.10 | 0.12 | 1.20 | 0.15 | 1.48 |
| 256 | 2.00 | 1.92 | 0.96 | 3.55 | 1.77 |
| 1024 | 41.53 | 55.52 | 1.34 | 78.68 | 1.89 |
| 4096 | 861.65 | 2118.92 | 2.46 | 1704.61 | 1.98 |

Table 4.2 Comparison in α

| n | PSTw | Peng | | Dijkstra | |
|---|---|---|---|---|---|
| | α | α | /.PSTw | α | /PSTw |
| 64 | 2.00 | 2.96 | 1.48 | 6.00 | 3.00 |
| 256 | 2.00 | 2.36 | 1.18 | 8.00 | 4.00 |
| 1024 | 2.01 | 2.48 | 1.23 | 10.00 | 4.98 |
| 4096 | 2.07 | 2.63 | 1.27 | 12.00 | 5.80 |

Fig. 4.2 and Table 4.2 show the comparison of $\alpha$. The /PSTw column shows the rate of $\alpha$ (called $\alpha$ rate) likewise. When $n$ = 4096, PSTw outperforms Peng by 1.27 times and Dijkstra by 5.8 times. The α rate of Dijkstra increases as $n$ increases, but it is difficult to judge whether that of Peng increases or not as $n$ increases from only the data. PSTw's α is about 2.0 and close to 1 although it slightly increases as $n$ increases. The PSTu's α′s are between 1.52 and in 1.71 in the same experimental situation [Yamane19] and we cannot have found out why PSTw's α is larger than PSTw's α respectively, and PSTw's α's are between 2.00 and 2.07 and it is slightly increasing.

To sum up, PSTw outperforms Peng and Dijkstra in both CPU time and $\alpha$, and especially Dijkstra greatly in $\alpha$.

## 4.2 Comparison in scale-free graphs

The result of comparison in sparse case ($n' = 2$) is mentioned in 4.2.1, and that in dense case ($n' = \sqrt{n}$) is mentioned in 4.2.2.

### 4.2.1 Sparse case ($n' = 2$)

Fig. 4.3 and Table 4.3 shows the comparison of the three algorithms in CPU time. In Fig. 4.3, PSTw's line overlaps with Dijkstra's. When $n = 4096$, PSTw is outperformed by Peng by 5.0 times and the CPU rate decreases as $n$ increases. On the other hand, PSTw and Dijkstra are almost the same and the CPU rate decreases as $n$ increases. Therefore, it seems that Dijkstra might slightly outperform PSTw when $n > 4096$.

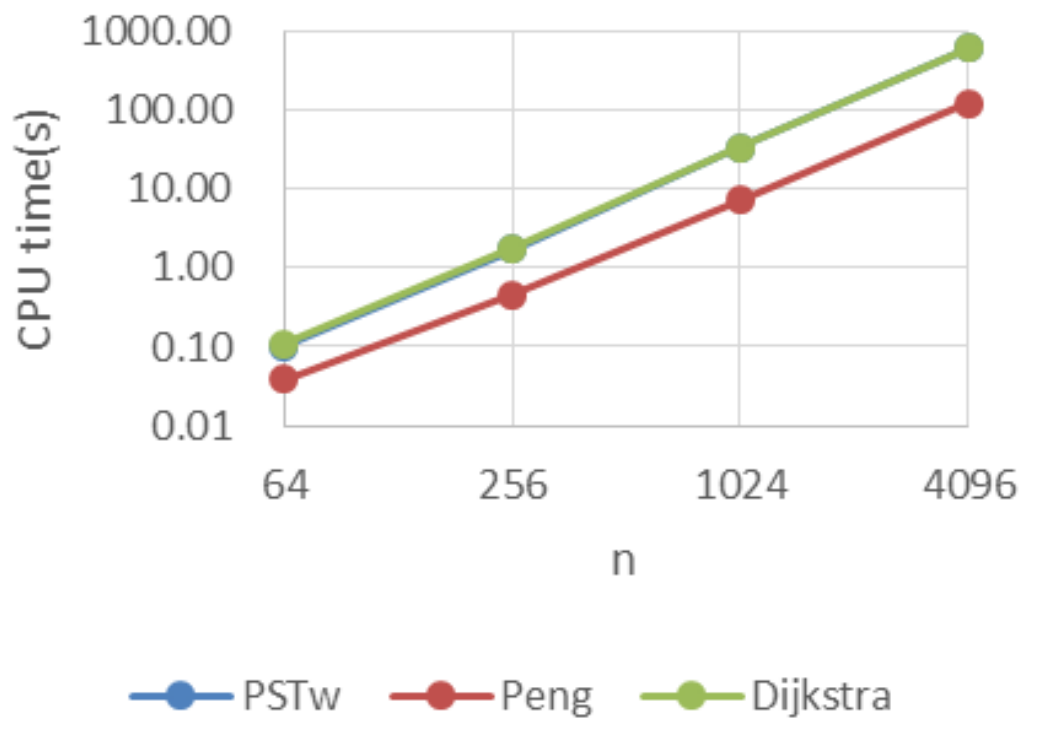

Fig. 4.3 Comparison in CPU time

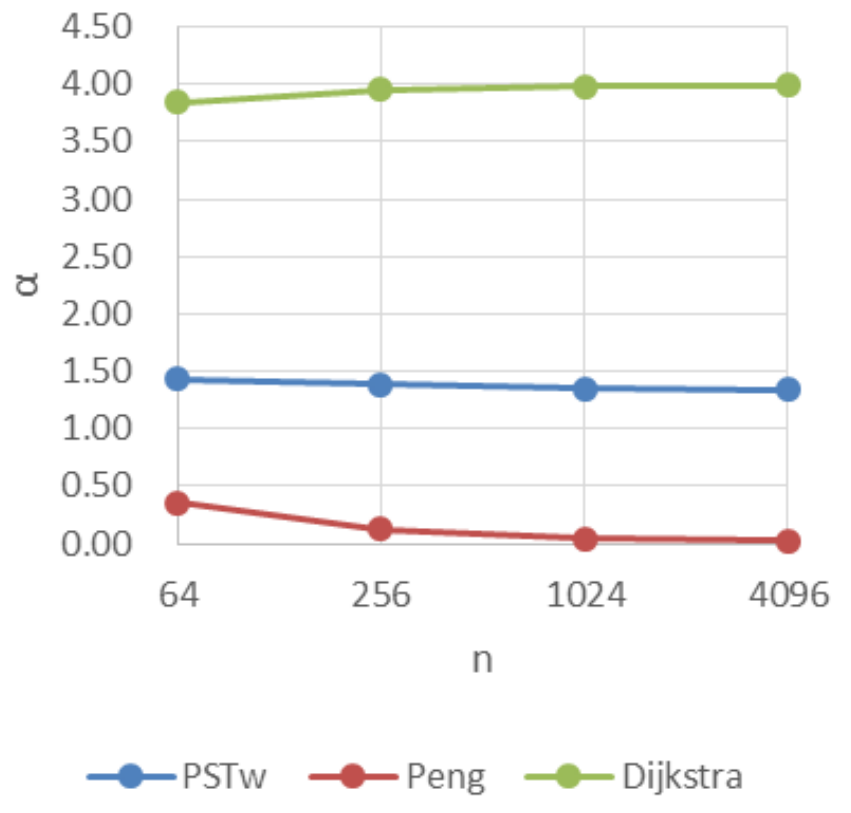

Fig. 4.4 Comparison in α

Table 4.3 Comparison in CPU time

| n | PSTw | Peng | | Dijkstra | |
|---|---|---|---|---|---|
| | time (s) | time (s) | /.PSTw | time (s) | /PSTw |
| 64 | 0.10 | 0.04 | 0.38 | 0.11 | 1.07 |
| 256 | 1.71 | 0.45 | 0.27 | 1.83 | 1.07 |
| 1024 | 33.75 | 7.32 | 0.22 | 33.94 | 1.01 |
| 4096 | 622.77 | 127.23 | 0.20 | 624.35 | 1.00 |

Table 4.4Comparison in α

| n | PSTw | Peng | | Dijkstra | |
|---|---|---|---|---|---|
| | α | α | /.PSTw | α | /PSTw |
| 64 | 1.43 | 0.36 | 0.25 | 3.84 | 2.69 |
| 256 | 1.39 | 0.13 | 0.09 | 3.95 | 2.84 |
| 1024 | 1.35 | 0.04 | 0.03 | 3.98 | 2.95 |
| 4096 | 1.34 | 0.02 | 0.01 | 3.99 | 2.98 |

Fig.4.4 and Table 4.4 shows the comparison in α. When $n = 4096$, PSTw is outperformed by Peng by 67 times. It is amazing that Peng's α is 0.02 because we thought α is not less than 1. In addition, the α rate even decreases as $n$ increases. On the other hand, PSTw outperforms Dijkstra by 3.0 times and the α rate increases as $n$ increases, but it seems that the α rate is approaching some upper bound.

PSTw's $\alpha$ is 1.34, a close value to 1 when $n = 4096$ and it decreases as $n$ increases.

To sum up, Peng outperforms PSTw and Dijkstra in CPU time and α and especially amazingly in α. Dijkstra slightly outperforms PSTw or they are the almost same on CPU time or they are the almost same because the CPU rate decreases and the α rate increases as $n$ increases. PSTw outperforms Dijkstra on α.

### 4.2.2 Dense case ($n' = \sqrt{n}$)

Fig.4.5 and Table 4.5 shows the comparison of CPU time. When $n = 4096$, PSTw outperforms Peng by 1.13 times and Dijkstra by 5.5 times, and the CPU time rate of both algorithms increases as $n$ increases. When $n \leq 1024$ Peng outperforms PSTw in CPU time, but PSTw slightly outperforms Peng when $n = 4096$, and the CPU rate of Peng increases as $n$ increases, so we judge that PSTw outperforms Peng in CPU time in dense scale-free graphs.

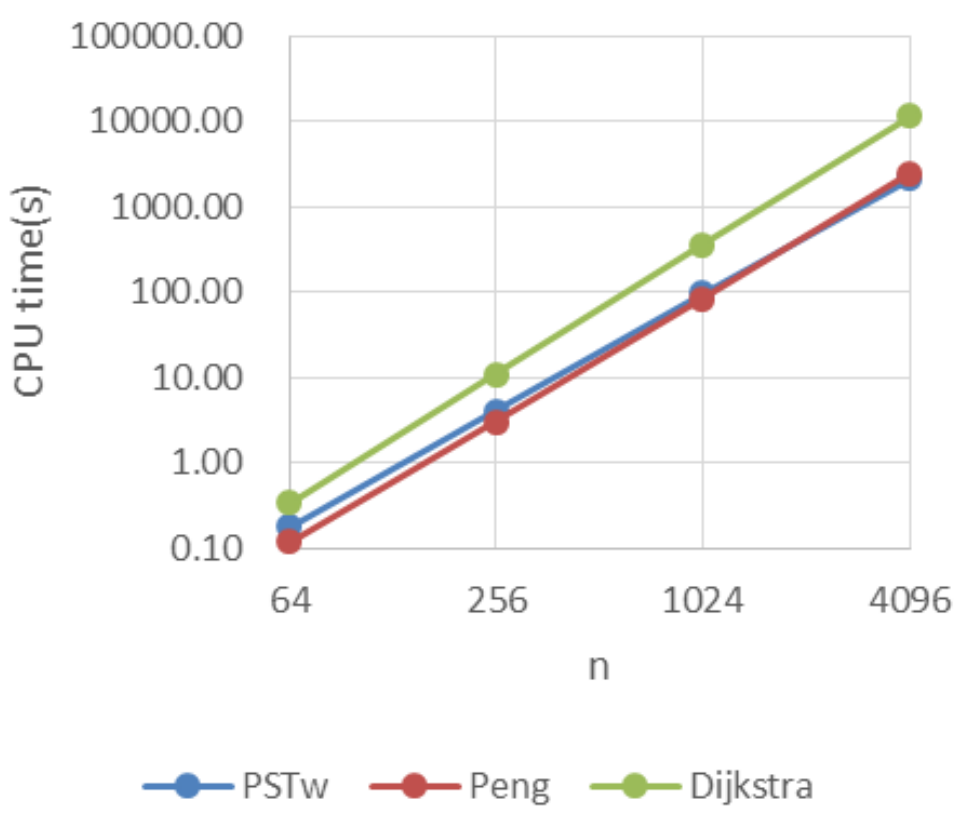

Fig. 4.5 Comparison in CPU time

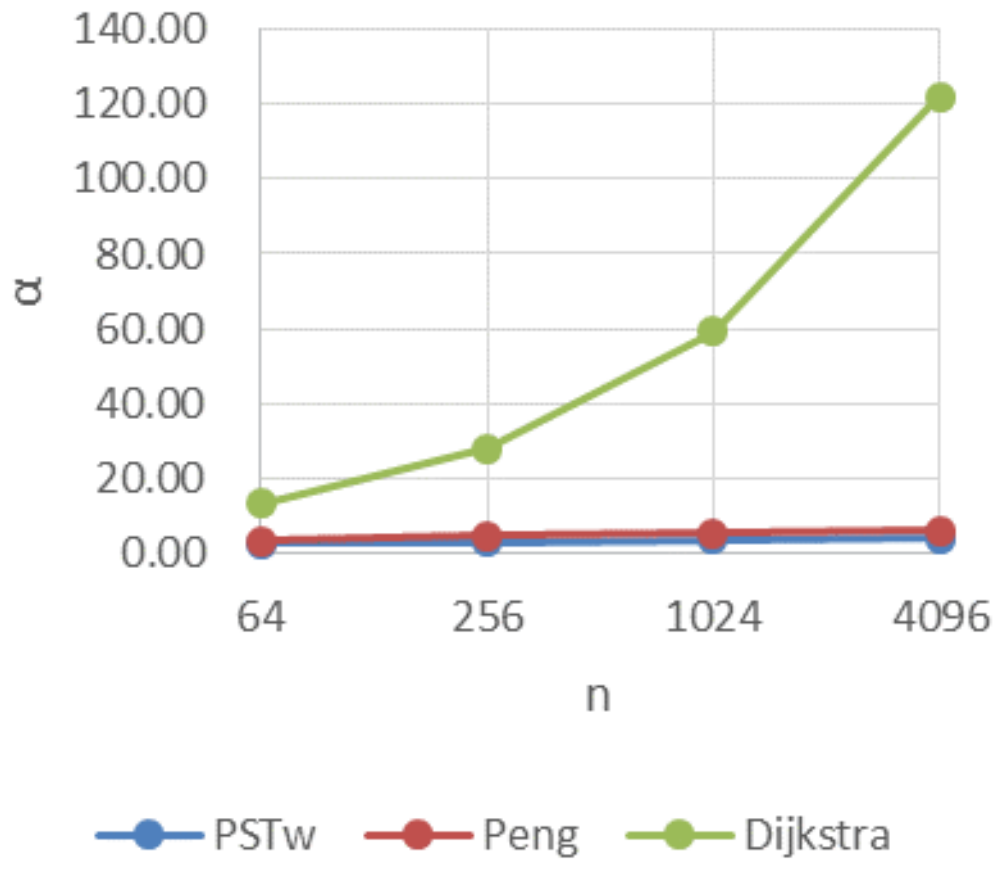

Fig. 4.6 Comparison in α

Table 4.5 Comparison in CPU time

| n | PSTw | Peng | | Dijkstra | |
|---|---|---|---|---|---|
| | time (s) | time (s) | /.PSTw | time (s) | /PSTw |
| 64 | 0.18 | 0.12 | 0.66 | 0.34 | 1.88 |
| 256 | 4.12 | 3.08 | 0.75 | 10.96 | 2.66 |
| 1024 | 97.99 | 83.27 | 0.85 | 361.91 | 3.69 |
| 4096 | 2172.52 | 2457.07 | 1.13 | 11621.44 | 5.35 |

Table 4.6 Comparison in α

| n | PSTw | Peng | | Dijkstra | |
|---|---|---|---|---|---|
| | α | α | /.PSTw | α | /PSTw |
| 64 | 2.65 | 3.42 | 1.29 | 13.22 | 4.99 |
| 256 | 3.16 | 4.84 | 1.53 | 28.22 | 8.93 |
| 1024 | 3.63 | 5.50 | 1.52 | 59.26 | 16.33 |
| 4096 | 4.20 | 6.14 | 1.46 | 122.08 | 29.07 |

Fig. 4.6 and Table 4.6 show the comparison of α. PSTw outperforms Peng by 1.5 times and Dijkstra by 29 times when $n = 4096$. The α rate of Dijkstra increases as $n$ increases, but it is difficult to judge whether the α rate of Peng increases or not as $n$ increases from only the data. The α rate of Dijkstra increases as $n$ increases.

PSTw's $\alpha$ is 4.2 when $n = 4096$ and it increases as $n$ increases. When the graph is dense, the depth of the shortest path trees tends to be small. Therefore, as mentioned above, it does not hold true that the value of $\alpha$ is close to 1, but it is sufficiently small in comparison with Dijkstra's value 122.

To sum up, PSTw outperforms Peng and Dijkstra in CPU time and $\alpha$, and especially Dijkstra in $\alpha$.

## 4.3 Consideration

We show our consideration on the result of comparison and space complexity.

### 4.3.1 On the result of comparison

Our consideration on 2) and 3) are very similar to ours on PSTu in [Yamane19], although ours on 1) is added newly.

1) Comparison with Peng

In hypercube-shaped and dense scale-free graphs, PSTw outperforms Peng in CPU time and $\alpha$. Conversely in sparse scale-free graphs, Peng outperforms PSTw greatly, and especially amazingly in $\alpha$. It is amazing that Peng's $\alpha$ is 0.02, and very close to 0 when $n = 4096$ because we thought $\alpha$ is not less than 1.

In Peng, when traversing a vertex $v$ whose shortest paths are all computed, pruning proceeds quickly by computing the distances to other vertices from $v$. And it computes all the shortest paths of the vertices whose degree is large earlier by sorting the vertices in the decreasing order of the degree of vertices. This mechanism works excellently in sparse scale-free graphs. This seems to be the reason why Peng outperforms PSTw greatly or amazingly. On the other hand, PSTw outperforms Peng in hypercube-shaped and dense scale-free graphs. The reason seems to be as follows: The degree of the vertices of a hypercube-shaped graph takes only one value, so the mechanism does not work so well. In dense scale-free graphs, there are a lot of vertices whose degree is large, that is, hubs, so the mechanism does not work so well in comparison with sparse ones.

2) Comparison with Dijkstra

PSTw outperforms Dijkstra in CPU time and $\alpha$ except that Dijkstra slightly outperforms PSTw or they are almost the same in case of sparse scale-free graphs. The reason seems as follows: Dijkstra's $\alpha$ is about equal to the average degree of each vertex. On the other hand, PSTw's $\alpha$ is close to 1 in case of hypercube-shaped and sparse scale-free graphs. In case of dense scale-free graphs, PSTw's $\alpha$ is 4.2 and not close to 1, but it is 29 times smaller than Dijkstra's $\alpha$ 122. In short, PSTw's $\alpha$ is much smaller than Dijkstra's $\alpha$, and that seems to be why PSTw outperforms Dijkstra.

3) On PSTw's $\alpha$

It is about 2.0 in hypercube-shaped graphs although it slightly increases as $n$ increases. It is also between 1.34 and 1.43 in sparse scale-free graphs and it also decreases as $n$ increases. These values are closes to 1. On the other hand, in case of dense scale-free graphs it is between 2.65 and 4.20, which are far from 1. The reason seems as follows: Let a vertex $w$ be adjacent to a source vertex $v$. Then $T(w)$ contains all the vertices adjacent to $w$, so they cannot be pruned from $T(v)$. The depth of $T(v)$ tends to be small in dense scale-free graphs. This seems to cause that $\alpha$ is far from 1. It even increases as $n$ increases.

### 4.3.2 On space complexity

Our consideration on space complexity is very similar to ours on PSTu in [Yamane19]. The relationship between PSTu and BSF corresponds to that between PSTw and Peng or Dijkstra. PSTw, Peng and Dijkstra needs $O(n^2)$ memory to store a distance matrix and a shortest path tree matrix. In addition, Dijkstra and Peng only needs $O(n+m)$ memory to represent graphs, and memory for a priority queue. However, in addition to the memory mentioned above, PSTw needs the memory to store $T(v)$ and a priority queue at each vertex $v$. Therefore, Dijkstra and Peng outperform PSTw from the viewpoint of space complexity.

## 5. Conclusion

We proposed a new all-pairs shortest path search algorithm for weighted graphs, which reduces $\alpha$ to the value close to 1 based on pruning by the shortest path trees of adjacent vertices of a source vertex when the depth of the shortest paths is relatively large. We also showed PSTw is deadlock-free; that is not a problem in PSTu. The results mentioned above show the following:

1) PSTw outperforms Peng in CPU time and $\alpha$ in hypercube-shaped and dense scale-free graphs, but conversely Peng outperforms PSTw greatly in CPU time and $\alpha$, and especially amazingly in $\alpha$.
2) PSTw outperforms Dijkstra in CPU time and $\alpha$ except that Dijkstra slightly outperforms PSTw or they are almost the same in case of sparse scale-free graphs.
3) Like PSTu, PSTw's $\alpha$ is close to 1 in case of hypercube-shaped and sparse scale-free graphs. In case of hypercube-shaped graphs it slightly increases, and in case of sparse scale-free graphs it decreases as $n$ increases like PSTu.
4) Like PSTu, in case of dense scale-free graphs, PSTw's $\alpha$ is between 2.65 and 4.20 and far from 1. The reason seems that the shortest paths cannot be pruned in the vertices adjacent to a source vertex, and the depth of the shortest path trees tends to small.
5) Like PSTu, Peng and Dijkstra outperform PSTw from the viewpoint of space complexity.

PSTw and PSTu are similar, and 2) to 5) show that although it is a little different that in 2) Dijkstra slightly outperforms PSTw or they are almost the same in sparse scale-free graphs.

As shown in this paper, PSTw and Peng have merits and demerits. We think it is necessary to compare in the other graphs and analyze time complexity to make them clearer.

## References

[BFS] Wikipedia's title: "Breadth-first search"

[Dijkstra] Wikipedia's title: Dijkstra's algorithm.

[Floyd62] R. W. Floyd. Algorithm 97: Shortest Path. CACM 5 (6): 345, 1962.

[Kim18] J. W. Kim, H. Choi, and S. Bae. Efficient Parallel All-Pairs ShortestPaths Algorithm for Complex Graph Analysis. Proceedings of International Conference on Parallel Processing Companion, 2018.

[Peng12] W. Peng, X. Hu, F. Zhao, and J. Su. A Fast algorithm to find all-pairs shortest paths in complex networks. Procedia Computational Science 9: 557-566, 2012.

[Warshall62] S. Warshall. A theorem on Boolean matrices. JACM 9 (1): 11–12, 1962.